\documentclass[twocolumn,aps,pra,showpacs,superscriptaddress,nofootinbib]{revtex4-1}

\usepackage[latin1]{inputenc}
\usepackage{amsmath}
\usepackage{graphicx,xcolor}
\usepackage{amssymb}
\usepackage{bm}
\usepackage{multirow}

\newcommand{\OO}{\mathcal P}
\newcommand{\ket}[1]{|#1\rangle}
\newcommand{\bra}[1]{\langle#1|}
\newcommand{\ident}{\leavevmode\hbox{\small1\normalsize\kern-.33em1}}

\begin{document}
\title{Experimental quantum process tomography of non trace-preserving maps}

\author{Irene Bongioanni}
\affiliation{Dipartimento di Fisica, Sapienza Universit\'a di Roma, I-00185 Roma, Italy}
\author{Linda Sansoni}
\affiliation{Dipartimento di Fisica, Sapienza Universit\'a di Roma, I-00185 Roma, Italy}
\author{Fabio Sciarrino}
\affiliation{Dipartimento di Fisica, Sapienza Universit\'a di Roma, I-00185 Roma, Italy}
\affiliation{Istituto Nazionale di Ottica (INO-CNR), L.go E. Fermi 6, I-50125 Firenze, Italy}
\author{Giuseppe Vallone}
\affiliation{Museo Storico della Fisica e Centro Studi e Ricerche Enrico Fermi, Via Panisperna 89/A, Compendio del Viminale, I-00184 Roma, Italy}
\affiliation{Dipartimento di Fisica, Sapienza Universit\'a di Roma, I-00185 Roma, Italy}
\author{Paolo Mataloni}
\affiliation{Dipartimento di Fisica, Sapienza Universit\'a di Roma, I-00185 Roma, Italy}
\affiliation{Istituto Nazionale di Ottica (INO-CNR), L.go E. Fermi 6, I-50125 Firenze, Italy}

\begin{abstract}
The ability of fully reconstructing quantum maps is a fundamental task of quantum information, 
in particular when coupling with the environment and experimental imperfections of devices 
are taken into account. In this context we carry out a quantum process tomography (QPT) approach for a set of non trace-preserving maps. We introduce an operator $\OO$ 
to characterize the state dependent probability of success for the process under investigation.
We also evaluate the result of approximating the process with a trace-preserving one.
\end{abstract}
\maketitle

\section{Introduction}
\begin{figure}[t]
 		\centering
		\includegraphics[width=7cm]{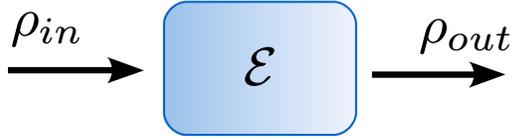}
		\caption{(Color online) Scheme of a generic quantum process $\mathcal{E}$.}
		\label{blackbox}
	\end{figure}

The complete characterization of quantum devices represents one of the fundamental tasks of quantum information science. The characterization of single- and two-qubit devices is particularly important since single-qubit and two-qubit controlled-NOT gates are the two building blocks of a quantum computer \cite{niel00bk}. Furthermore, identifying an unknown quantum process acting on a quantum system is another key task for quantum dynamics control, in particular in presence of decoherence \cite{kofm09pra,mohs08pra}. In this context any quantum process $\mathcal{E}$ can be described by a linear map \cite{niel00bk} 
acting on density matrices $\rho$ associated to a Hilbert space $\mathcal{H}$ which transforms an input state $\rho_{in}$ into an output state $\rho_{out}$ (Fig. \ref{blackbox}):
\begin{equation}
	\rho_{in}\stackrel{\mathcal{E}}{\longrightarrow}\rho_{out}=\mathcal{E}(\rho_{in}).
\end{equation}

The complete characterization of such a process can be realized by reconstructing the corresponding map $\mathcal E$. The action induced by a black box may be represented by a \textit{process matrix} $\chi$ which is experimentally reconstructed by quantum process tomography (QPT) \cite{chua97jmo,dari03prl,mohs08pra,rohd05pra,mohs06prl,whit07jsb}.
So far, several QPT experiments have been performed for trace-preserving processes, such as single-qubit transmission channels \cite{alte03prl,howa06njp}, optimal transpose map \cite{scia04pra},
gates for ensembles of two-qubit
systems in NMR \cite{chil01pra}, a two-qubit quantum-state filter \cite{mitc03prl}, a universal two-qubit gate \cite{poya97prl}, Control-NOT (CNOT) and Control-Z (CZ) gates for photons \cite{obri04prl,lang05prl,kies05prl}.

Recently theoretical and experimental analyses of non trace-preserving processes have been carried out. Kiesel \textit{et al.} evaluated the role of prior knowledge of the intrinsic feature of the experimental setup in order to obtain physical and easily understandable parameters for characterizing the gate and estimating its performance \cite{kies05prl}.
Furthermore quantum process tomography in presence of decoherence has been analyzed for a fast identification of the main decoherence mechanisms associated to an experimental map \cite{kofm09pra}.

Here we address the characterization of non trace-preserving maps, 
focusing on the evaluation of an operator $\OO$, representing the success probability of the process.
In particular we carry out a quantum process tomography (QPT) approach for a set of non trace-preserving maps. Then, we discuss possible errors occurring in presence of inappropriate approximations.

The paper is organized as follows. In Section II a brief review of the main theoretical aspects of QPT and of the process fidelity, both for trace-preserving and non trace-preserving maps, is presented. In Section III we report an example of QPT of a non trace-preserving process, corresponding to the transformation induced by a partially transmitting polarizing beam splitter. The QPT experimental realization and results are then presented together with a brief discussion on possible wrong approaches to the problem, when a non trace-preserving process is approximated with a trace-preserving one. Finally, 
the conclusions are given in Section IV.

\section{Quantum process tomography}
Consider an unknown quantum process, i.e. a black box, acting on a physical quantum system described by a density matrix $\rho$ associated to a $d$-dimensional Hilbert space $\mathcal{H}$. A complete characterization of the process may be obtained by the Kraus representation of quantum operations in an open system \cite{chua97jmo,poya97prl,mohs08pra}.
A generic map $\mathcal{E}$ acting on a generic state $\rho$ can be expressed by the Kraus representation: \cite{niel00bk}
\begin{equation}
	\mathcal{E}(\rho)=\sum_i{E_i\rho E_i^{\dagger}}
	\label{E}
\end{equation}
where $E_i$ are operators acting on the system and satisfying the relation\footnote{$\sum_i{E_i^{\dagger}E_i}\leq\mathbb{I}$ means that
the eigenvalues of the hermitian operator $\sum_i{E_i^{\dagger}E_i}-\mathbb{I}$ are not positive.}
 {$\sum_i{E_i^{\dagger}E_i}\leq\mathbb{I}$}. 
If $\mathcal{E}$ is a trace-preserving process 
the completeness relation {$\sum_i{E_i^{\dagger}E_i}=\mathbb{I}$} holds.

The quantum process tomography of $\mathcal E$ consists of the experimental reconstruction 
of the operators $\left\{E_i\right\}$. In order to relate each operator $E_i$ with measurable parameters it is convenient to use a \textit{fixed} basis of operators $\left\{A_i\right\}$ such that:
\begin{equation}
	E_i=\sum_m{a_{im}A_m}.
	\label{A}
\end{equation}
By substituting this expression in (\ref{E}), the map reads as follows
\begin{equation}
	\mathcal{E}(\rho)=\sum_{m,n}{\chi_{mn}A_m\rho A_n^{\dagger}}\;,
	\label{E2}
\end{equation}
where $\chi_{mn}=\sum_i{a_{im}a^{\star}_{in}}$. {By construction, the matrix $\chi_{\mathcal E}$
with elements $\chi_{mn}$ is hermitian and semidefinite positive.}

To experimentally reconstruct each element $\chi_{mn}$ we prepare $d^2$ input states $\rho_k$ forming a basis for the Hilbert space of $d\times d$ matrices. The output states can be written as
\begin{equation}
	\mathcal{E}(\rho_k)=\sum_j{\lambda_{kj}\rho_j},
\end{equation}
where the coefficients $\lambda_{kj}$ are experimentally obtained by characterizing $\mathcal{E}(\rho_k)$ 
and expressing it in the $\left\{\rho_k\right\}$ basis.
{By defining the coefficients $\beta^{mn}_{jk}$ such that}
\begin{equation}
	A_m\rho_jA_n^{\dagger}=\sum_k{\beta_{jk}^{mn}\rho_k}\;,
\end{equation}
{it is easy to obtain a relation} between $\lambda_{kj}$ and $\chi_{mn}$ \cite{chua97jmo}:
\begin{equation}
	\sum_{m,n}{\beta_{jk}^{mn}\chi_{mn}}=\lambda_{jk}\;.
\end{equation}
In order to find the matrix $\chi_{\mathcal E}$ which completely describes the process $\mathcal{E}$, we need to operate a matrix inversion of $\beta_{jk}^{mn}$. If $\tau_{jk}^{mn}$ is this generalized inverse matrix
({i.e. $\sum_{jk}\tau^{pq}_{jk}\beta^{mn}_{jk}=\delta_{pm}\delta_{qn}$}), 
the elements of $\chi_{\mathcal E}$ read:
\begin{equation}
	\chi_{mn}=\sum_{jk}\tau_{jk}^{mn}\lambda_{jk}.
\end{equation}

For a non trace-preserving map, it is important to consider not only 
the transformation acting on a generic input state, but also the probability of success of the map.
For a given input state $\rho$, the probability of obtaining an output state from the black box is given by
\begin{equation}
	\text{Tr}\left[\mathcal{E}(\rho)\right]=\text{Tr}\left[\sum_{mn}{\chi_{mn}A_m\rho A_n^{\dagger}}\right]=\text{Tr}\left[\OO\rho\right]\;,
\end{equation}
where $\OO$ is a semidefinite positive hermitian operator defined as:
\begin{equation}
	\OO=\sum_{mn}\chi_{mn}A^{\dagger}_nA_m\leq\mathbb{I}\;.
	\label{Pt}
\end{equation}
Let's write $\OO$ in its diagonal form, $\OO=\sum_i{p_i\ket{p_i}\bra{p_i}}$, where $\ket{p_i}$ are the eigenstates and $0\leq p_i\leq1$ the corresponding eigenvalues.
Different cases may occur:
\begin{itemize}
\item[i)] $p_i=1\ \forall i$, i.e. $\OO=\mathbb{I}$ for a trace-preserving process; 
\item[ii)] $p_i=p<1\ \forall i$ ($\OO$ is proportional to the identity operator) for a non trace-preserving process with state independent success probability;
\item[iii)]  there is at least one eigenvalue $p_i$ different from the others in the case of a non trace-preserving process with state dependent success probability. 
\end{itemize}
The eigenvectors of $\OO$ coincide with the {``probability of success''} 
eigenstates of the transformation.\\

We now describe how to compare two quantum processes. 
It is well known that a quantum state can be completely determined by a tomographic reconstruction \cite{jame01pra} and compared with the expected theoretical state by a variety of measures, such as
quantum state fidelity \cite{jozs94jmo}.
Similarly, we know that the process matrix $\chi_{\mathcal E}$ gives a convenient way of representing
a general operation $\mathcal{E}$. A closely related but more abstract representation
is provided by the Jamiolkowski isomorphism \cite{jami72rmp}, which relates a quantum operation $\mathcal{E}$ to a quantum state, $\rho_{\mathcal{E}}$:
\begin{equation}
	\rho_{\mathcal{E}}\equiv\left(\mathbb{I}\otimes\mathcal{E}\right)\ket{\Phi}\bra{\Phi},
\end{equation}
where $\ket{\Phi}=\frac{1}{\sqrt{d}}\sum_{j}{\ket{j}\ket{j}}$ is a maximally entangled state associated to the \textit{d-dimensional} system with another copy of itself, and $\left\{\ket{j}\right\}$ is an orthonormal basis set.
If $\mathcal E$ is a trace-preserving process, then the quantum state $\rho_{\mathcal E}$ is
normalized, $\text{Tr}[\rho_{\mathcal E}]=1$.
 In this way, by associating a quantum process to a quantum state, for two trace-preserving processes $\mathcal{E}$ and $\mathcal{G}$, a 
\textit{Process Fidelity} $\Delta$ has been defined as follows \cite{ragi01pla,niel02pla,gilc05pra,wang06pra}
\begin{equation}
	\Delta(\mathcal{E},\mathcal{G})=\mathcal{F}(\rho_{\mathcal{E}},\rho_{\mathcal{G}})
	\label{Fpro}
\end{equation}
where $\mathcal{F}$ is the quantum state fidelity $\mathcal{F}=\text{Tr}\left[\sqrt{\sqrt{\rho_{\mathcal{E}}}\rho_{\mathcal{G}}
\sqrt{\rho_{\mathcal{E}}}}\right]^2$ \cite{jozs94jmo}. It is easy to demonstrate that, by choosing 
the set $A_m=\left\{\sqrt{d}\ket{i}\bra{j}\right\}$ as Kraus operators, we have $\rho_{\mathcal{E}}\equiv\chi_{\mathcal{E}}$, and, in general, $\mathcal{F}(\rho_{\mathcal{E}},\rho_{\mathcal{G}})=\mathcal{F}(\chi_{\mathcal{E}},\chi_{\mathcal{G}})$ if 
any complete set of operators $A'_m$ satisfying $\text{Tr}[A'_mA'^\dag_n]=d\delta_{mn}$ is used
($\delta_{mn}$ is the Kronecker delta).
Thus, if we want to compare an experimental map $\chi$ with the expected one $\chi_{id}$, 
the process fidelity is 
\begin{equation}
	\Delta=\text{Tr}\left[\sqrt{\sqrt{\chi}\ \chi_{id}\sqrt{\chi}}\right]^2.
	\label{fidProc}
\end{equation}
The last expression gives the fidelity of density matrices with unit trace. 
However, if $\chi$ represents a \textit{non trace-preserving} process, i.e. 
{$\text{Tr}[\chi]=\frac{1}{d}\text{Tr}[\OO]<1$}, 
the process fidelity definition is generalized as follows \cite{kies05prl}. 
Let $\chi_{id}$ be the ideal matrix associated to a non trace-preserving process 
in the Kraus representation and $\chi$ the experimental one. 
The fidelity for such a process is written as
\begin{equation}
	\Delta(\chi,\chi_{id})=\frac{\text{Tr}\left[\sqrt{\sqrt{\chi}\ \chi_{id}\sqrt{\chi}}\right]^2}{\text{Tr}\left[\chi\right]\text{Tr}\left[\chi_{id}\right]}.
	\label{FidNTP}
\end{equation}
Note that the physical meaning of this expression is the same of (\ref{fidProc}): indeed we can express it as
\begin{equation}
	\Delta(\chi,\chi_{id})=\Delta(\chi^{\prime},\chi_{id}^{\prime})=\text{Tr}\left[\sqrt{\sqrt{\chi^{\prime}}\ \chi_{id}^{\prime}\sqrt{\chi^{\prime}}}\right]^2
\end{equation}
where $\chi^{\prime}=\frac{\chi}{\text{Tr}[\chi]}$ and $\chi^{\prime}_{id}=\frac{\chi_{id}}{\text{Tr}[\chi_{id}]}$ are well defined physical states ($\text{Tr}[\chi^{\prime}]=\text{Tr}[\chi^{\prime}_{id}]=1$) which, however, 
do not correspond to any meaningful quantum operation, since the probability of success 
of the corresponding processes will be larger than 1 for some input states
(i.e. the corresponding operators $\OO$ will have at least one eigenvalue larger than 1).

It is interesting to highlight that the process fidelity defined in (\ref{FidNTP}) does not distinguish between two processes $\mathcal{E}$ and $\mathcal{G}$ if $\mathcal{E}=\alpha\mathcal{G}$, where $\alpha$ is a constant, i.e. two processes are indistinguishable if they differ only for a global loss, as it often occurs in the experimental implementations of photonic quantum systems.

\section{QPT of a partially transmitting polarizing beam splitter}
Now we analyze a simple example of the quantum process tomography of a non trace-preserving, 
state dependent map,
acting on a single polarization qubit ($d=2$). 
Consider a partially transmitting polarizing beam splitter (PPBS) with trasmittivities $T_H$ and $T_V$ 
at the horizontal and vertical polarization, respectively. Following the Kraus approach, in which we consider $A_j$ as the Pauli operators, we report the analytical expression of the process matrix $\chi_{\text{PPBS}}$.

In general, if we inject a photon with arbitrary polarization state into the PPBS 
the output state will be:
\begin{equation}
	\alpha\ket{H}+\beta\ket{V}\rightarrow\alpha\sqrt{T_H}\ket{H}+\beta\sqrt{T_V}\ket{V},
	\label{trasf}
\end{equation}where $\alpha,\beta\in\mathbb{C}$ and $|\alpha|^2+|\beta|^2=1$.
Clearly the probability of success of this transformation is state dependent. Let's write the process matrix associated to this map. According to QPT calculations, we fix the Pauli matrices $\sigma_i,\ i=0,...,3$, in a bidimensional Hilbert space as the basis $A_i$  in the Kraus sum 
(satisfying the normalization condition $\text{Tr}[A_mA^\dag_n]=d\delta_{mn}$), 
and choose the set $\left\{\rho_k\right\}$ of the states to be measured, obtaining the following matrix:
\begin{equation}
\chi_{\text{PPBS}}=
{	\begin{pmatrix}
		\frac{(\sqrt{T_H}+\sqrt{T_V})^2}{4}& &0& &0& &\frac{T_H-T_V}{4}\\
		0& &0& &0& &0\\
		0& &0& &0& &0\\
		\frac{T_H-T_V}{4}& &0& &0& &\frac{(\sqrt{T_H}-\sqrt{T_V})^2}{4}
	\end{pmatrix}}.
	\label{chimat}
\end{equation}
Obviously, the explicit form of $\chi_{\text{PPBS}}$ does not depend on the chosen set $\left\{\rho_k\right\}$, 
but only on the {fixed basis $A_i$ in the Kraus representation}.
{Let's now  write the explicit form of the operator $\OO$ for the PPBS.}
By using the $\chi$ matrix given in \eqref{chimat}, we obtain
\begin{equation}
\OO_{\text{PPBS}}=
\begin{pmatrix}
	&T_H& &0&\\
	&0& &T_V&
\end{pmatrix}.
\label{P}
\end{equation}
This operator is proportional to the identity only when $T_H=T_V$.

\subsection{Experimental QPT of a PPBS}
\begin{figure}[ht]
		%\centering
		\includegraphics[width=0.8\columnwidth]{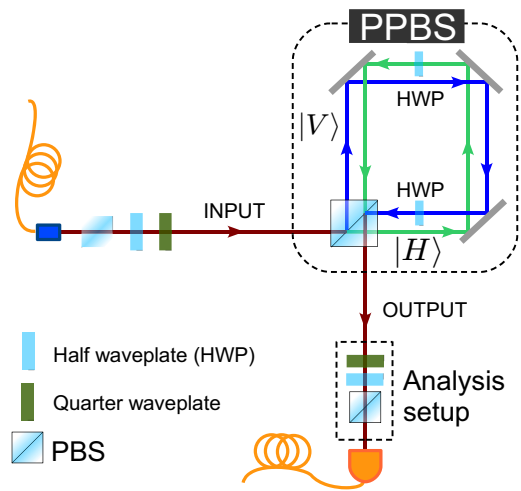}
		\caption{(Color online) Experimental setup used for the QPT of a partially transmitting polarizing beam splitter. The PPBS is implemented by a displaced Sagnac interferometer and two half-waveplates. The measurements are performed with a standard polarization analysis setup.}
		\label{setup}
	\end{figure}

In this subsection we report the experimental realization of QPT for a partially transmitting polarizing beam splitter. In the experimental setup shown in Fig. \ref{setup} the PPBS is implemented by a closed-loop scheme,
{similar to the one used in \cite{okam09sci,tomo07sci}},
operating with two half-waveplates (HWP). A diagonally polarized light beam is splitted by a polarizing beam splitter (PBS) in two beams with equal intensity and orthogonal polarizations. Precisely, the horizontal ($H$) and vertical ($V$) components travel along two parallel directions inside the interferometer, counterclockwise and clockwise, respectively. One half-waveplate intercepts the $H$ beam, 
while the other intercepts the $V$ beam; by rotating the waveplates it is possible to vary the value of $T_V$ with respect to $T_H$.

The photons injected in this interferometric setup are generated by a spontaneous paramentric down conversion source realized with a nonlinear crystal cut for type II non collinear phase matching \cite{kwia95prl}. The crystal is pumped by a CW diode laser and pairs of degenerate photons are produced at wavelength $\lambda=806nm$. One of the photon is used as a trigger, while the other is delivered to the PPBS setup.
\begin{figure}[hb]
		\centering
		\includegraphics[width=0.95\columnwidth]{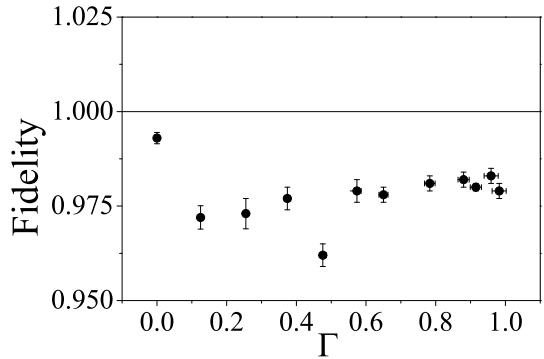}
		\caption{(Color online) Measurement of fidelity as a function of $\Gamma=\frac{T_H}{T_V}$.}
		\label{fidexp}
\end{figure}
\begin{figure}[ht]
	\centering
	\includegraphics[width=0.9\columnwidth]{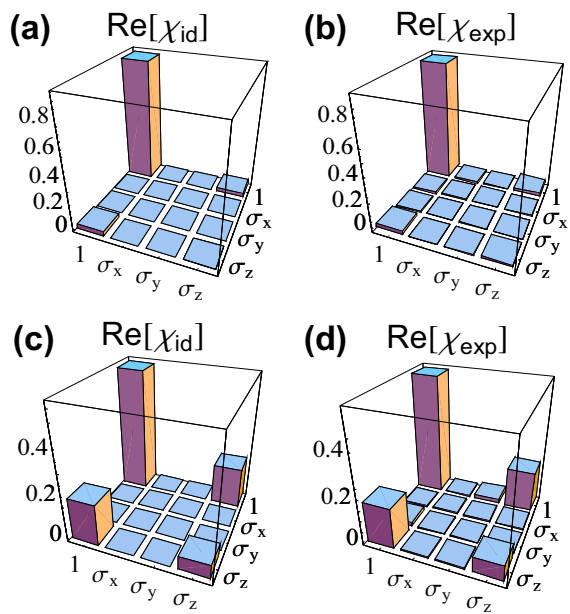}
	\caption{(Color online) Real part of ideal and experimental process matrices, $\chi_{id}$ and $\chi_{exp}$, for (\textit{a-b}) $\Gamma=0.879$ and (\textit{c-d}) $\Gamma=0.255$. The imaginary part are negligible.}
\label{chiML}
\end{figure}
We prepared six different input states, $\ket{H}$, $\ket{V}$, $\ket{D}$, $\ket{A}$, $\ket{R}$, $\ket{L}$ associated to horizontal, vertical, diagonal, anti-diagonal, right-handed and left-handed polarization respectively, and measured the six output components for each input with a standard polarization analysis setup.
We repeated this procedure for different values of the ratio $\Gamma=T_V/T_H$ and, for each value of $\Gamma$,
we reconstructed the experimental $\chi$ matrix of the process. We then performed an optimization of the process matrix following a \textit{maximum likelihood} approach \cite{jeze03pra,obri04prl}; in particular we minimized the following function
\begin{eqnarray}
\begin{aligned}
	f(\vec{t})=
\sum_{a,b=1}^{d^2}\frac{1}{n_{ab}}\Biggl[ n_{ab}-&\sum_{m,n=0}^{d^2-1}\bra{\psi_b}\hat{\sigma}_m\ket{\phi_a}\times\Biggr.\\
&\Biggl.\bra{\phi_a}\hat{\sigma}_n\ket{\psi_b}\tilde{\chi}_{mn}(\vec{t})\Biggr]^2
\label{lhood}
\end{aligned}
\end{eqnarray}
where $n_{ab}$ are the measured coincidence counts for the $a$th input and the $b$th output, $\ket{\phi_a}$ and $\ket{\psi_b}$ indicates the input and the output state respectively, and $\hat{\sigma}_m$ are the Pauli operators.
Since we are not interested into overall losses affecting the transformation
(even the adopted fidelity is independent of global losses) 
we normalize the experimental $\chi_{exp}$ matrix such that the maximum eigenvalue of $\OO$ is 1.
\begin{figure}[ht]
		\centering
		\includegraphics[width=0.9\columnwidth]{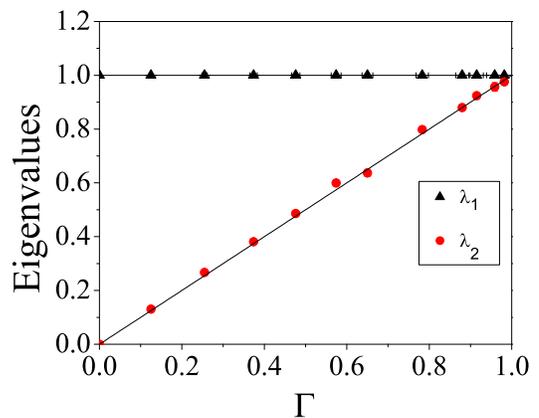}
		\caption{(Color online) Probability operator eigenvalues as a function of the ratio $\Gamma=T_V/T_H$. Solid lines represent expected behaviour.}
		\label{Peigenv}
	\end{figure}
We determined the fidelity between the experimental map and the ideal one for several values of
$\Gamma$, as shown in Fig. \ref{fidexp}. We observe that the process fidelity approaches
unity for each value of $\Gamma$, and in general, we have $F>96\%$
with a good agreement between the experimental data and the theory. 
In Fig. \ref{chiML} two examples of ideal and experimental process matrices, corresponding to
$\Gamma=0.879$ and $\Gamma=0.255$, are shown. 

We also estimated the probability operator $\OO$: the behaviour of its eigenvalues $\lambda_1$ and $\lambda_2$ as a function of $\Gamma$ is shown in Fig. \ref{Peigenv}.
We observe that $\lambda_1=1$ for each value of $\Gamma$ (by construction), while the other eigenvalue, $\lambda_2$, shows a decreasing behaviour as the ratio between the trasmittivities decreases, as expected from (\ref{P}).
Again, a very good agreement between experimental data and theory is obtained.

\subsection{Trace-preserving approximation}
\begin{figure}[t]
		\centering
		\includegraphics[width=0.9\columnwidth]{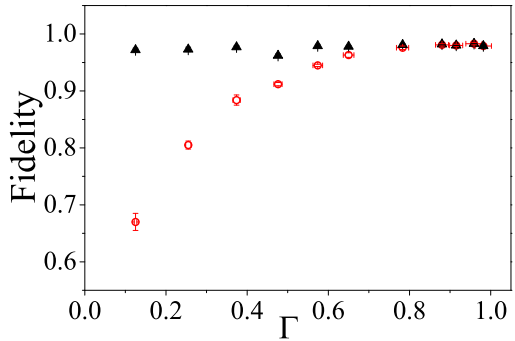}
		\caption{(Color online) Experimental fidelities calculated imposing the constraint $\sum_{m,n}{\chi_{mn}\sigma_n\sigma_m=\mathbb{I}}$ (red open circles). Fidelities obtained with the previous method are also reported (black filled triangles).}
		\label{fidvinc}	
\end{figure}

The method above described can be usefully adopted even when the process under investigation is ideally trace-preserving. In fact, when a quantum process tomography is practically implemented, any
interaction with the environment as well as experimental imperfections may cause the process to be non trace-preserving.
In practice, to approximate the process as a trace-preserving one corresponds to minimize the likelihood
function (\ref{lhood}) with the additional constraint $\OO=\sum_{m,n}{\chi_{mn}\sigma_n\sigma_m=\mathbb{I}}$.
In this way we are imposing the probability of success to be independent of the input state. 
We carried out the $f(t)$ minimization by taking into account
the constraint\footnote{We used the function {\tt NMinimize[\{f, cons\}, t]} of the MATHEMATICA\copyright5 program that allows to numerically minimizes $f(t)$ subject to the constraints {\tt cons}.
$\mathcal P=\mathbb I$.
Note that the constraint imposes the normalization Tr$[\chi_{exp}]=1$.}
and evaluated the process fidelity between the obtained $\chi_{exp}$ 
and the ideal matrix (\ref{chimat}) 
for each value of $\Gamma$. The results are shown in Fig. \ref{fidvinc}.
\begin{figure}[t]
		\centering
		\includegraphics[width=0.9\columnwidth]{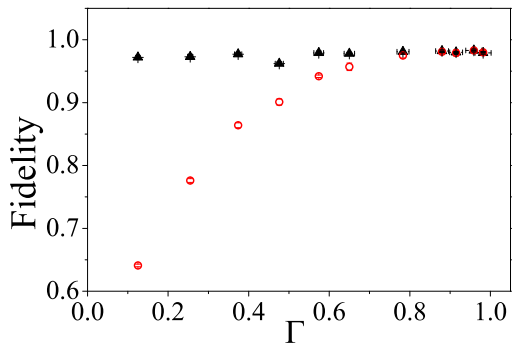}
		\caption{(Color online)   Experimental fidelities calculated using the post-selective approach (red open circles). Fidelities obtained with the correct method are also reported (black filled triangles).}
		\label{fidnorm}	
\end{figure}
As expected, this method gives results similar to those obtained in Section IIIA for 
$\Gamma\sim 1$, while the fidelities values are different as $\Gamma$ decreases. 
In particular, the fidelities calculated by imposing the constraint decrease as $\Gamma$ goes to zero. 
It is evident that constraining the process to be trace-preserving does not allow
to correctly reconstruct the associated map.

{A further scenario where probability of success must be taken into account
may arise when measurements are performed in post-selection.
The reconstruction of the output state density matrices (which obviously are normalized physical states) 
for several input state, leads to a trace-preserving process.}
Even in this case we evaluated the fidelities between the resulting process matrix and the ideal one obtaining the results shown in Fig. \ref{fidnorm}. As in the previous case the fidelity decreases as 
$\Gamma$ goes to zero.
Note that this approach is not correct even from a theoretical point of view: 
the process matrix $\chi_{\mathcal E}$ 
obtained by normalizing the output states {could be non-physical
(i.e. it could have negative eigenvalues)} 
and its expression depends on the chosen set of input states. This is due to the fact that normalization implies the process to be no longer a linear map and equation (\ref{E2}) is not valid anymore.
In general, the output state normalization produces wrong process matrices
for {\it any} non trace-preserving operation with state dependent success probability.

\section{conclusions}

A review on quantum process tomography of non trace-preserving maps has been reported. The experimental implementation of a simple non trace-preserving, state dependent process, i.e. the transformation induced by a partially polarizing beam splitter, provided process fidelities larger than $96\%$ for any value of the ratio between the transmittivities $\Gamma$.
Particular attention has been addressed to the state dependence property of the process through 
evaluation of the operator $\OO$ (\ref{Pt}). This operator has been calculated and measured in the case of a PPBS and its eigenvalues resulted to be different from unity (see (\ref{P})), as expected for a non trace-preserving process.
In order to stress the validity of the method a brief discussion about possible wrong approaches 
has been presented together with the explicit calculation of the PPBS process fidelities. 
The obtained results clearly show that approximation of a non trace-preserving, state dependent process with a trace-preserving one does not allow a correct reconstruction of the real process map.

{QPT of non trace preserving processes are relevant for
linear optical logic gates with success probability $<1$. 
Indeed, typically it is just assumed that the success probability of such gates is uniform across
input states and hence it is crucial to check the validity of this assumption 
for any application. 
For example, it can be interesting to investigate whether losses in the planar 
integrated waveguide chips currently being used \cite{poli09sci} could affect different input states differently.
}

\acknowledgements
This work was supported by Finanziamento Ateneo 2009 of Sapienza Universit\`a di Roma.

%\bibliography{Bibliografia}
%\bibliographystyle{prl_with_titles}

\end{document}